\documentclass[aps,pre,twocolumn,floatfix,superscriptaddress,showpacs,showkeys]{revtex4-2}
\usepackage{epsf,amsmath,amssymb,verbatim,color,multirow,pifont,mathrsfs,soul,amsthm, wasysym}
\usepackage{graphicx, upgreek}
\usepackage[us,12hr]{datetime}

\begin{document}

\title{Discontinuous percolation via suppression of neighboring clusters in a network}
\author{Young Sul Cho}
\email{yscho@jbnu.ac.kr}
\affiliation{Department of Physics, Jeonbuk National University, Jeonju 54896, Rep. of Korea}
\affiliation{Research Institute of Physics and Chemistry, Jeonbuk National University, Jeonju 54896, Rep. of Korea}

\date{\today}

\begin{abstract}
Our recent study on the Bethe lattice reported that a discontinuous percolation transition emerges as the number of occupied links increases and each node rewires its links to locally suppress the growth of neighboring clusters. 
However, since the Bethe lattice is a tree, a macroscopic cluster forms as an infinite spanning tree but does not contain a finite fraction of the nodes. 
In this paper, we study a bipartite network that can be regarded as a locally tree-like structure with long-range neighbors. In this network, each node in one of the two partitions is allowed to rewire its links to nodes in the other partition to suppress the growth of neighboring clusters. We observe a discontinuous percolation transition characterized by the emergence of a single macroscopic cluster containing a finite fraction of nodes, followed by critical behavior of the cluster size distribution. We also provide an analytical explanation of the underlying mechanism.
\end{abstract}

\maketitle

\section{Introduction}

An explosive percolation (EP) transition refers to the phenomenon that the probability of a node belonging to a macroscopic cluster increases abruptly as the fraction of occupied links exceeds a threshold~\cite{Achlioptas:2009, explosive_phenomena, souza_nphy, ziff_ncomm, universalEP2021, yschosciadv2025}. EP can be observed in various real systems~\cite{makse, chopre:2011, nanotube, fracture, crackling, experiment}.
To reveal this phenomenon theoretically, each link is occupied in a way that suppresses the formation of large clusters, using information of cluster sizes.
Specifically, the optimal link that minimizes the resulting cluster size is selected from a set of candidate links and then occupied. If the number of candidate links is finite, this is referred to as local cluster size information; otherwise, it is referred to as global cluster size information. When local information of cluster sizes is used, the transition is abrupt but continuous~\cite{ziff_lattice:2010,filippo_pre:2010,fss_exp,local,tricritical,hklee,choi,dacosta_prl,dacosta_pre,grassberger,riordan,smoh:2016,eppre:2023,epprl:2023, Jan_gap1,Jan_gap2, smohgrowing_2021}. In contrast, when global cluster size information is used, a discontinuous transition occurs~\cite{local2018, choprl:2016, park_hybrid, choi_hybrid, Half_restricted, cho_science, chopre:2010, largest, gaussian, resource, smohdicpt_2018}.

In the field of complex systems, a central focus is the emergence of macroscopic phenomena driven by entities interacting with their neighbors using only local information, such as $k$-core percolation~\cite{kcore, Bootstrap, universalHPT}, percolation in interdependent networks~\cite{InterdependentHavlin, InterdependentDorogovtsev} and in networks with higher-order interactions~\cite{triadic2023}, and explosive synchronization~\cite{explosivesync2011, chkim2025, slee_ncomm2025}.

Following this line of inquiry, our previous study~\cite{yschosciadv2025} introduced a mechanism in which
each node rewires its links to the optimal neighboring candidates that minimize the resulting cluster size, also based on local cluster size information,
analogous to how quarantine strategies change depending on the local infection situation~\cite{quaratine}. This approach led to the observation of a rigorous discontinuous percolation transition (DPT) on a Bethe lattice.
However, since the Bethe lattice is a tree, multiple macroscopic clusters form as infinite spanning trees at the threshold.
Consequently, such a macroscopic cluster does not contain a finite fraction of the nodes, despite the fact that in many other percolation models, a macroscopic cluster typically includes a finite fraction of the nodes~\cite{stauffer, er, kim_percolation}.
We are interested in whether a DPT with a single macroscopic cluster---similar to EP models on random networks~\cite{Achlioptas:2009}---can arise through suppression based on local cluster size information.

A macroscopic cluster in a random network near the threshold typically exhibits a locally tree-like structure with long-range neighbors. Motivated by this observation, we consider here
a locally tree-like network in which the formation of large clusters is suppressed via link rewiring. Specifically, only one end of each link is allowed to rewire, so that the number of occupied links for each rewiring node is conserved and analytical results can be derived. To this end, we consider a bipartite network in which only the nodes in one of the two partitions are allowed to rewire their links.

We begin with a bipartite network with constant degree. In this network, each node in one of the two partitions rewires its occupied links to neighbors within the other partition, in ascending order of their cluster sizes, repeatedly until the largest cluster size reaches a steady state. We show that a DPT, characterized by the formation of a single macroscopic cluster, emerges once the fraction of occupied links exceeds a threshold. Finally, we numerically observe a DPT in bipartite networks with both Poisson and scale-free degree distributions.

This paper is organized as follows. In Sec.~\ref{sec:model}, we introduce the model used in this study, and in Sec.~\ref{sec:discpt}, we observe a DPT of a single macroscopic cluster and provide an analytical explanation of this phenomenon. Then in Sec.~\ref{sec:cdist} we examine the critical behavior of the cluster size distribution following the DPT. In Sec.~\ref{sec:discussion}, we conclude by discussing the implications of our findings. Appendix A provides a supplemental analytical analysis to explore the origin of the critical behavior observed in the cluster size distribution. In Appendix B, we numerically observe a DPT in bipartite networks with both Poisson and scale-free degree distributions, and in Appendix C, we summarize the comparison between the current and previous results~\cite{yschosciadv2025}.

\section{Model}
\label{sec:model}

\begin{figure}[t!]
\includegraphics[width=0.7\linewidth]{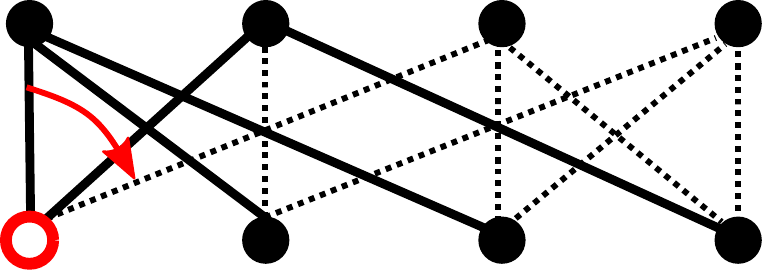}
\caption{Bipartite network with $z=3$ and $N=8$ divided into two partitions, represented as the top and bottom rows.
Solid lines represent occupied links, while dotted lines indicate unoccupied links.
A node ($\bigcirc$) with $n=2$ occupied links is randomly selected from the bottom row.
After the node disconnects all of its links, its three neighbors (from left to right in the top row) belong to clusters of sizes three, two, and one, respectively.
To occupy $n$ links in ascending order of neighbor cluster sizes, the node occupies links to the two rightmost neighbors.
As a result of this process, the location of the occupied link changes as indicated by the arrow.  
}
\label{Fig:Bipartite_diagram_single}
\end{figure}

We construct a bipartite network of size $N$ consisting of two partitions of nodes, indexed by $1\leq i \leq N/2$ and $N/2+1 \leq i \leq N$. Each node is connected to $z$ randomly selected neighbors in the opposite partition via undirected links.
Initially, each link is occupied with a given probability $p$. A cluster is defined as a set of nodes that are connected through occupied links.

At each time step, a node is randomly selected from the first partition ($1 \leq i \leq N/2$), 
where the number of occupied links attached to the node is denoted by $n$ ($0 \leq n \leq z$).
The selected node rewires its occupied links according to the {\it suppression rule} through the following steps:
\begin{itemize}
\item[(i)] Disconnect all occupied links. 
\item[(ii)] Evaluate the cluster sizes of all $z$ neighbors. 
\item[(iii)] Occupy $n$ links, choosing neighbors in ascending order of their cluster sizes. 
\end{itemize}
An example of this rewiring process is illustrated in Fig.~\ref{Fig:Bipartite_diagram_single}.

The rewiring process is repeated until the size of the largest cluster reaches a steady state.
Various measurements are then performed at this steady state as functions of $p$.

\section{Discontinuous percolation of a single macroscopic cluster}
\label{sec:discpt}

\begin{figure}[t!]
\includegraphics[width=1.0\linewidth]{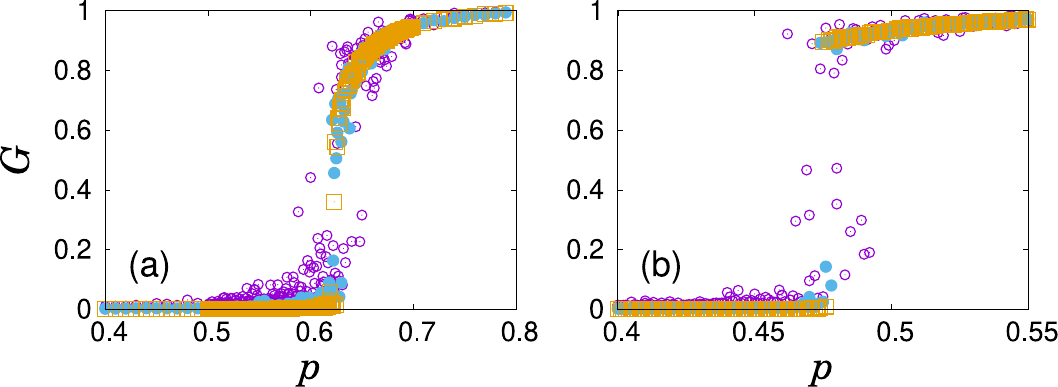}
\caption{$G(p)$ for a single realization with $N/10^3 = 1$ ($\circ$), $8$ ($\bullet$), and $64$ ($\square$), shown for (a) $z = 3$ and (b) $z = 4$.}
\label{Fig:Bipartite_Rewire_G}
\end{figure}

We measure the order parameter $G$, defined as the fraction of nodes belonging to the largest cluster, as a function of $p$.
We observe that $G$ exhibits a discontinuous jump to $G_0>0$ at the threshold, described by
\begin{equation}
G(p) = \left\{\begin{array}{lll}
0 & \textrm{~for~} & p < p_c \\\\
G_0 & \textrm{~at~} & p = p_c
\end{array}\right.
\label{Eq:Pinf}
\end{equation}
and increases gradually for $p \geq p_c$ when $z=3$ and $4$, as shown in Fig.~\ref{Fig:Bipartite_Rewire_G}(a) and (b), respectively. 
We note that the order parameter $G$ indicates the presence of a macroscopic cluster comprising a finite fraction of nodes, 
in contrast to the order parameter used in our previous work~\cite{yschosciadv2025}, 
which measured the probability that a node belongs to a macroscopic cluster.

To analytically understand the origin of the phenomenon, we introduce two probabilities, $P_{\infty}$ and $Q_{\infty}$, 
defined at the steady state. Here, $P_{\infty}$ represents the probability that a node in the first partition (the set of nodes allowed to rewire their links) belongs to a macroscopic cluster, while $Q_{\infty}$ denotes the probability that a node in the second partition (the set of nodes not allowed to rewire their links) belongs to a macroscopic cluster. Then, $G=(P_{\infty}+Q_{\infty})/2$ when there is only one macroscopic cluster.

It is difficult to derive exact formulas for $Q_{\infty}$ and $P_{\infty}$, as the probability that a node belongs to a macroscopic cluster varies depending on whether its neighbor currently belongs to a macroscopic cluster or not. Moreover, a node in the first partition may rewire its links, potentially disconnecting it from a macroscopic cluster and thus affecting its neighbors in turn.

For an analytical interpretation of the origin of the DPT, we assume that a node in the first partition belongs to a macroscopic cluster independently at random with probability $P_{\infty}$. Then we have
\begin{equation}
Q_{\infty}(p)=1-(1-pP_{\infty}(p))^z,
\label{eq:QP_relation}
\end{equation} 
where we use the property that a node in the second partition belongs to a macroscopic cluster only if it is connected via an occupied link (occupied with probability $p$ independently at random) to at least one neighbor in a macroscopic cluster. Note that the suppression rule is not applied to nodes in the second partition.

We derive the self-consistency equation for $P_{\infty}$ by applying the suppression rule to nodes in the first partition, yielding
\begin{align}
&P_{\infty}(p)=\notag \\
             &\sum_{n=1}^z\binom{z}{n}p^n(1-p)^{z-n}\bigg[\sum_{m=z-n+1}^z\binom{z}{m} R_{\infty}^m(p)(1-R_{\infty}(p))^{z-m}\bigg],
\label{eq:Pselfconsist}
\end{align}
where $R_{\infty}=1-(1-pP_{\infty})^{z-1}$ represents the probability that a neighbor of a node in the first partition is connected to a macroscopic cluster via its outgoing links. The term
$\binom{z}{n}p^n(1-p)^{z-n}$ represents the probability that a node in the first partition has $n$ occupied links, since each
link is initially occupied with probability $p$, and the number of occupied links for each node in the first partition remains unchanged during rewiring.
The summation within the large parentheses thus indicates that at least $z-n+1$ neighbors of the node must belong to a macroscopic cluster via their outgoing links in order for the node to be connected
to a macroscopic cluster using its $n$ occupied links.

\begin{figure}[t!]
\includegraphics[width=0.9\linewidth]{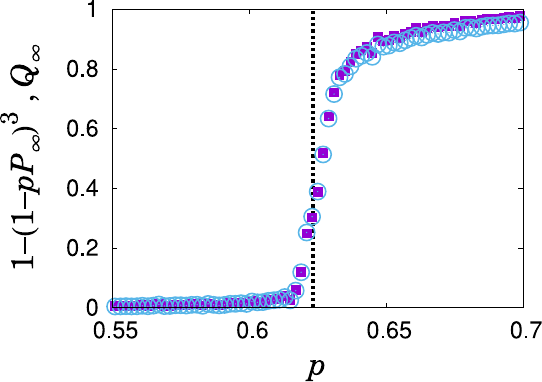}
\caption{Comparison of $1-(1-pP_{\infty}(p))^3(\bigcirc)$ and $Q_{\infty}(p)(\blacksquare)$ from a single realization for each $p$, with $N=16000$ and $z=3$. To obtain $P_{\infty}$ and $Q_{\infty}$,
we computed the fraction of nodes belonging to the largest cluster in the first and second partitions, respectively, at steady state. The vertical dotted line indicates $p_c$.}
\label{Fig:Bipartite_Rewire_QP}
\end{figure}

\begin{figure}[t!]
\includegraphics[width=0.9\linewidth]{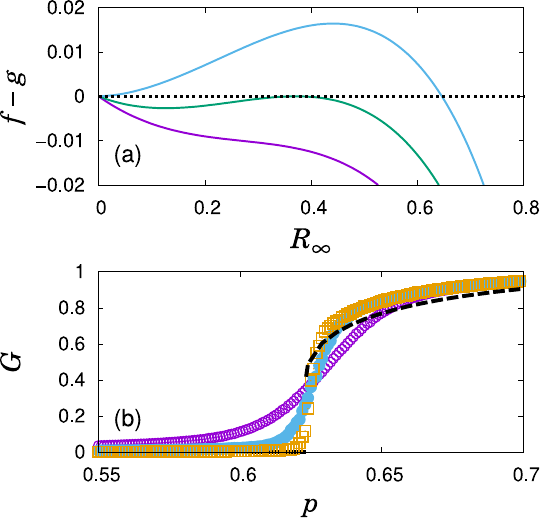}
\caption{(a) Plots of $f(R_{\infty}) - g(R_{\infty})$ for $p = 0.61$, $p = p_c$, and $p = 0.64$, from bottom to top, with $z = 3$.
(b) $G(p)$ for $N/10^3 = 1$ ($\circ$), $8$ ($\bullet$), and $64$ ($\square$) with $z = 3$.
The dashed line represents the theoretical curve that describes the behavior of $G(p)$ near $p_c$.}
\label{Fig:Bipartite_G_Theory}
\end{figure}

We apply Eqs.~(\ref{eq:QP_relation}) and (\ref{eq:Pselfconsist}) to clarify the origin of the DPT for $z=3$.
Eq.~(\ref{eq:QP_relation}) shows excellent agreement with the data up to $p_c$, as illustrated in Fig.~\ref{Fig:Bipartite_Rewire_QP}, where the theoretical value of $p_c$ is presented in the next paragraph.
This result suggests that the discontinuity in $G$ at $p_c$ can be explained within this theoretical framework.
We note that as $p$ increases beyond $p_c$, a discrepancy emerges between the two sides of the equation, likely due to increasing correlations among neighbors of a common node, which violate the assumptions underlying the equation.

By substituting $z=3$ into the right-hand side of Eq.~(\ref{eq:Pselfconsist}) and using the relation $(1-\sqrt{1-R_{\infty}})/p=P_{\infty}$, we derive the equation $f(R_{\infty})-g(R_{\infty})=0$, where $f(x)=3p^2(1-p)^2x^3+3p^3(1-p)(3x^2-2x^3)+p^4(3x-3x^2+x^3)$ and $g(x)=1-\sqrt{1-x}$. As shown in Fig.~\ref{Fig:Bipartite_G_Theory}(a), the equation has only the trivial solution $R_{\infty}=0$ for $p < p_c$. A nontrivial solution $R_{\infty}(p_c)>0$ emerges discontinuously at $p=p_c$, and $R_{\infty}(p)$ increases gradually as $p$ increases beyond this point. To determine $p_c$ and $R_{\infty}(p_c)$, we solve the equations $f(x)=g(x)$ and $f'(x)=g'(x)$ simultaneously, with $p=p_c$ and $x=R_{\infty}(p_c)$. This yields $p_c \approx 0.623339$ and $R_{\infty}(p_c) \approx 0.371835$.

We obtain $Q_{\infty}=1-(1-R_{\infty})^{3/2}$ and $P_{\infty}=(1-\sqrt{1-R_{\infty}})/p$ using the numerically calculated nontrivial solution $R_{\infty}>0$ of the equation $f(R_{\infty})=g(R_{\infty})$. Assuming the presence of a single macroscopic cluster, we use the relation $G = (P_{\infty}+Q_{\infty})/2$. The theoretical curve for $G$ obtained in this manner agrees well with the simulation results near $p_c$, as shown in Fig.~\ref{Fig:Bipartite_G_Theory}(b), and explains the origin of the DPT associated with the emergence of a macroscopic cluster. As a result, the estimated jump size of $G$ is $G_0 = (P_{\infty}(p_c)+Q_{\infty}(p_c))/2 \approx 0.417456$.

Although we specifically used $z=3$, we expect that a DPT occurs for $z \geq 3$ generally, as supported by the data for $z=4$ shown in Fig.~\ref{Fig:Bipartite_Rewire_G}(b). The origin of such DPT for general $z \geq 3$ could likewise be explained within this theoretical framework, using Eqs.~(\ref{eq:QP_relation}) and (\ref{eq:Pselfconsist}).

\section{Critical behavior of the cluster size distribution}
\label{sec:cdist}

We are interested in the evolution of the cluster size distribution $n_s(p)$, defined as the number of clusters of size $s$ over $N$.
To understand the dynamics, we derive an asymptotic equation for $n_s$ that represents the fraction of nodes belonging to clusters of size $s$. 
In this derivation, we use $P_s$ and $Q_s$ to denote the fraction of nodes belonging to clusters of size $s$ in the first and second partitions, respectively,
such that $sn_s = (P_s + Q_s)/2$.

\begin{figure}[t!]
\includegraphics[width=0.9\linewidth]{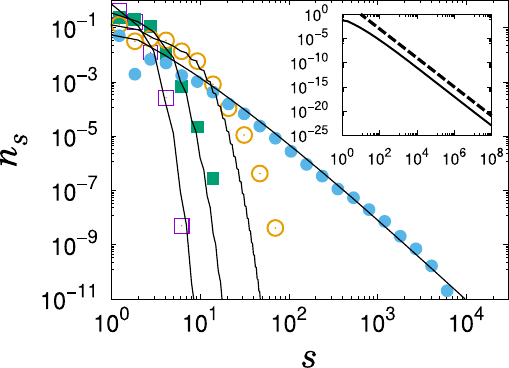}
\caption{Plots of $n_s$ for $p=0.1 (\square)$, $0.3 (\blacksquare)$, $0.5 (\bigcirc)$, and $0.63 (\bullet)$ with $N=16000$.
The solid lines represent the theoretical curves for $p = 0.1$, $0.3$, $0.5$, and $p_c$, from left to right. We note that the curve for $n_s$ at $p = 0.63>p_c$ aligns well with the theoretical prediction for $p_c$. This is due to finite-size effects, whereby the value of $p$ at which $n_s$ exhibits a power-law behavior decreases to $p_c$ as $N$ increases~\cite{PREhybrid2D}. Inset: Theoretical curve for $p=p_c$ (solid line), extended over a wide range of $s$.
The slope of the dashed line is $-3$, confirming the estimated critical exponent $\tau=3$.}
\label{Fig:Bipartite_Rewire_cdist}
\end{figure}

Appendix A presents asymptotic equations for $P_s$ and $Q_s$, derived under the above assumptions with the constraints $\sum_s P_s = 1-P_{\infty}$ and $\sum_s Q_s = 1-Q_{\infty}$. 
We numerically solve these equations and verify that $n_s = (P_s+Q_s)/2s$ agrees quite well with the simulation data, as shown in Fig.~\ref{Fig:Bipartite_Rewire_cdist}, supporting the validity of the equation in describing the evolution of $n_s$. Notably, $n_s$ shows the critical behavior $n_s \propto s^{-\tau}$ at $p_c$, which supports that the percolation transition is actually a hybrid phase transition~\cite{InterdependentHavlin, Bootstrap, kcore, BootstrapDorogovtsev, kcore3, universalHPT}.

To estimate the theoretical value of $\tau$, we approximate $n_s(p_c)=As^{-\tau}$ for $s \geq s_c$, where $A$ is an $s$-independent constant and $s_c$ is an arbitrary large positive integer. For sufficiently large $s_c$, the normalization condition $\sum_s sn_s(p_c)=1-G_0$ can be approximated as
\begin{equation}
1-G_0 = A\int_{s_c}^{\infty} s^{1-\tau}ds + \sum_{s=1}^{s_c}sn_s(p_c).
\label{eq:norm}
\end{equation}
By connecting Eq.~(\ref{eq:norm}) with $A=s_c^{\tau}n_{s_c}(p_c)$,
we estimate $\tau$ as $\tau = 2 + s_c^2n_{s_c}(p_c)/(1-G_0-\sum_{s=1}^{s_c}sn_s(p_c))$, where both $s_c^2n_{s_c}(p_c)$ and $\sum_{s=1}^{s_c}sn_s(p_c)$ are obtained by numerically solving the equations for $n_s(p_c)$ with a sufficiently large $s_c$. We find that the estimated value of $\tau$ approaches $\tau=3$ from below as $s_c$ increases. Based on this behavior, we conclude that the theoretical value is $\tau = 3$ (see the inset of Fig.~\ref{Fig:Bipartite_Rewire_cdist}).

\section{Discussion}
\label{sec:discussion}

In this paper, we studied a bipartite network of degree $z=3$ in which each link is initially occupied with probability $p$.
Each node in the first partition repeatedly rewires its occupied links to neighboring clusters in ascending order of their cluster size until the largest cluster size reaches a steady state. As $p$ exceeds the critical point $p_c$, 
a macroscopic cluster containing a finite fraction of the nodes emerges discontinuously.
We expect that this phenomenon generally occurs for $z \geq 3$, and in general networks that can be regarded as locally tree-like structures, such as Erd\H{o}s--R\'enyi networks~\cite{er}. 
To demonstrate the generality of this result, in Appendix B we numerically observe a DPT in bipartite networks with Poisson and scale-free degree distributions.

This result is significant because, unlike previous EP models, it shows that the discontinuous emergence of a macroscopic cluster containing a finite fraction of nodes can occur from a suppression rule applied among a finite set of clusters.
In contrast, in the previous EP models each link is occupied sequentially by selecting the best option to suppress among a set of candidate clusters.
From this dynamical perspective, the modification in this study is that a link is occupied randomly as $p$ increases by $p \rightarrow p+1/N$,
and links are rewired according to the suppression rule until the system reaches a steady state.
In this manner, a link is added to the steady state.

A possible rewiring process leading to a steady state after each link addition is described as follows. When a link is added to the initial rewiring node, its cluster size increases. Nodes connected to the rewiring node via occupied links then sequentially rewire their links, and they are not allowed to rewire further. In this manner, rewiring nodes spread outward from the initial node. This spread of rewiring terminates when no additional nodes can rewire. 

\begin{figure}[t!]
\includegraphics[width=0.9\linewidth]{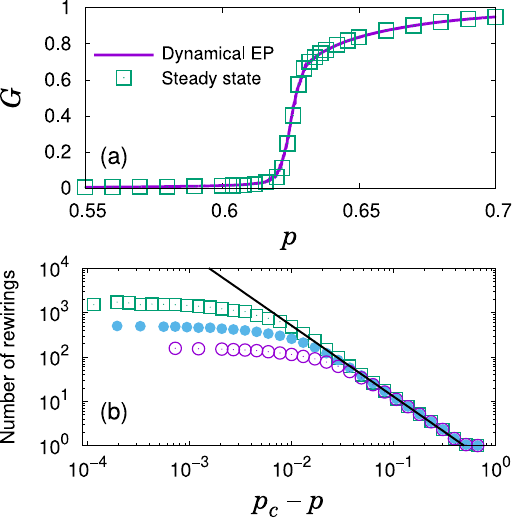}
\caption{(a) $G$ from the dynamical EP process (line) and the steady state ($\square$) coincide on the bipartite network with $z=3$ and $N/10^3=16$. (b) Number of rewiring nodes after each link occupation in the dynamical EP process for $N/10^3=1$ $(\circ)$, $4$ $(\bullet)$, and $16$ $(\square)$.
The slope of the solid line is $-8/5$.}
\label{Fig:Bipartite_G_nRewire_EP}
\end{figure}

Such a spread of rewiring from the initial node is sufficient to reach a steady state in a tree network. We numerically confirmed that a steady state is achieved through this process, as shown in Fig.~\ref{Fig:Bipartite_G_nRewire_EP}(a), indicating that the network effectively behaves as a tree within the spread size. Moreover, the spread size remains finite before the threshold but diverges as $(p_c-p)^{-8/5}$ as the threshold is approached from below, as shown in Fig.~\ref{Fig:Bipartite_G_nRewire_EP}(b). This result implies that our dynamical model exhibits a DPT based on the sizes of a finite number of clusters before the threshold, whereas traditional explosive percolation requires information on an infinite number of cluster sizes to produce a DPT even before the threshold. A more detailed clarification of the underlying dynamics will be presented in future work.

A bipartite network was used in this study, while in our previous work~\cite{yschosciadv2025} we employed a Bethe lattice to demonstrate that a DPT can emerge through a suppression rule implemented via link rewiring.
We note that both networks can be effectively analyzed using the locally tree-like property, which enables us to examine how collaborative suppression of macroscopic cluster formation leads to a DPT. For a comparison of the two results, see Appendix C.
It would be interesting future work to investigate whether the same phenomenon occurs in non–tree-like networks, such as square lattices.

\section*{Acknowledgments}

This research was supported by Global - Learning \& Academic research institution for Master’s·PhD students, and Postdocs(LAMP) Program 
of the National Research Foundation of Korea(NRF) grant funded by the Ministry of Education(No. RS-2024-00443714), and by NRF grant No. RS-2025-16067164.

\section*{Appendix A: analytical analysis of the cluster size distribution}

In this Appendix, we provide an asymptotic equation for $sn_s = (P_s+Q_s)/2$ when $z=3$.
We define $R_s$ as the probability that a node in the second partition---reached by following a randomly chosen link from a neighbor in the first partition---belongs to a cluster of size $s$. The size $s$ is determined by the node's two outgoing neighbors, as described below. 
We note that a node in the second partition is not subject to the suppression rule. Therefore, each of its two outgoing links is occupied with probability $p$, and the neighbor it connects to belongs to a cluster of size $s$ with probability $P_s$.

It is noteworthy that, in contrast to the case of random bond percolation in a bipartite network~\cite{newmanpre2001}, the sum of the sizes of the clusters reachable through occupied outgoing links is not taken into account when deriving $n_s$. This is because occupied outgoing links may be rewired repeatedly, and thus the assumption that each neighbor independently belongs to a cluster of size $s$ with probability $P_s$ must be employed to reflect the steady state behavior.

\begin{figure}[t!]
\includegraphics[width=1.0\linewidth]{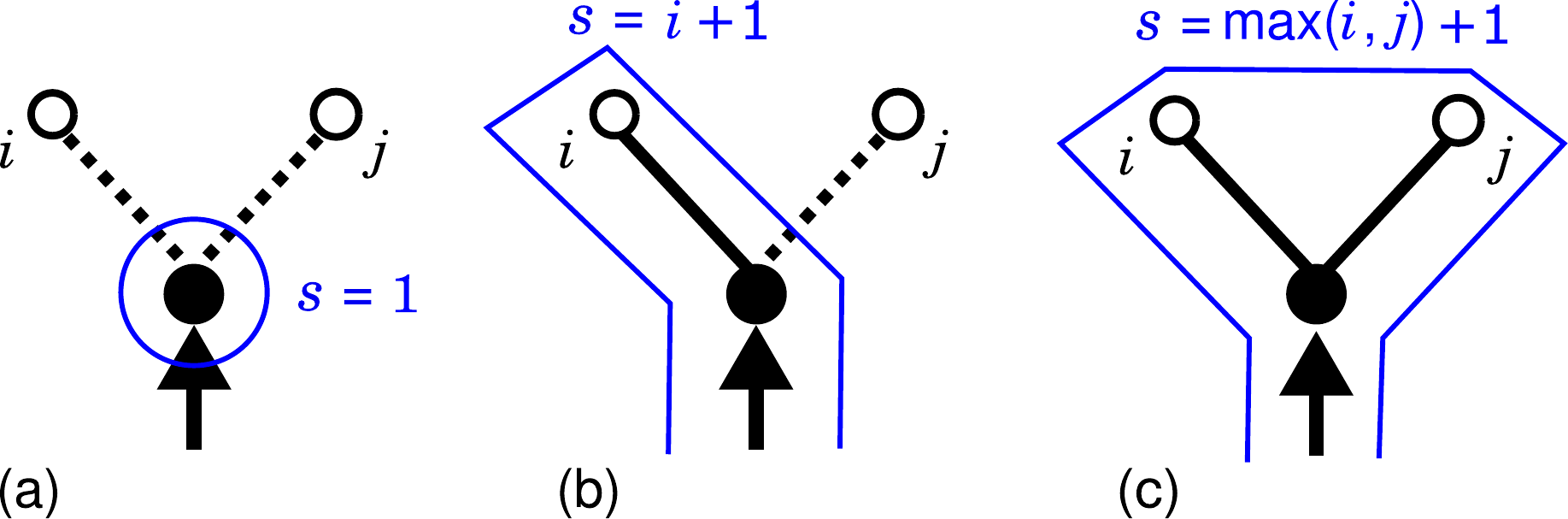}
\caption{Illustration of the determination of the cluster size $s$ of a node $(\bullet)$ for computing $R_s$ in the cases where (a) both outgoing links are unoccupied, (b) only one outgoing link is occupied, and (c) both outgoing links are occupied. In all three cases, the cluster sizes of outgoing neighbors are initially assigned random values, denoted by $i$ and $j$.
We note that $i$, $j$, and $s$ do not represent the partial cluster sizes in the outgoing directions from the ingoing arrow, and thus the relation $s=i+j+1$ does not hold.}
\label{Fig:Rs_schematic}
\end{figure}

Depending on the occupancy status of the outgoing links, we consider the following three cases, as illustrated in Fig.~\ref{Fig:Rs_schematic}.
If both outgoing links are unoccupied, we let $s=1$ [Fig.~\ref{Fig:Rs_schematic}(a)]. 
If one outgoing link is occupied and the other is unoccupied, then the node belongs to a cluster of size $s$ with probability $P_s$ satisfying $s=i+1$, where $i$ is the randomly assigned size of the cluster reached through the occupied link [Fig.~\ref{Fig:Rs_schematic}(b)] and 1 is added to account for the node itself. When both outgoing links are occupied and the sizes of the clusters reached via these links are randomly assigned to $i$ and $j$, then $i$ and $j$ may differ. This contradicts the property that all nodes connected through occupied links belong to the same cluster. To resolve this contradiction, we assume that the node and the two neighbors reached via the occupied links all belong to the same cluster whose size is given by $s=\textrm {max}(i,j)+1$. Therefore, for the node to belong to a cluster of size $s$ including the two outgoing neighbors, the larger of the two cluster sizes must be $s-1$ [Fig.~\ref{Fig:Rs_schematic}(c)].

By accounting for these cases in the derivation of $R_s$, we obtain
\begin{equation}
R_s = (1-p)^2\delta_{s1}+2p(1-p)P_{s-1}+p^2\bigg(P_{s-1}^2+2P_{s-1}\sum_{i=1}^{s-2}P_i\bigg),
\tag{A1}
\label{Eq:Rs}
\end{equation}
where the three terms on the right-hand side correspond to the three cases discussed above, in order.
This equation satisfies $\sum_s R_s = 1-(1-pP_{\infty})^2$.

Similarly, we derive $Q_s$ by assuming that a node in the second partition is connected to $0 \leq n \leq 3$ neighbors with probability $\binom{3}{n}p^n(1-p)^{3-n}$, and that its cluster size $s$ is given by 1 plus the maximum of the $n$ connected cluster sizes, where each size $i$ is independently drawn according to the distribution $P_i$. This gives 
\begin{align}
Q_s=&(1-p)^3\delta_{s1}+3p(1-p)^2 P_{s-1} \notag \\
       &+3p^2(1-p)\bigg(P_{s-1}^2+2P_{s-1}\sum_{i=1}^{s-2}P_i\bigg) \notag \\
       &+p^3\bigg(P_{s-1}^3+3P_{s-1}^2\sum_{i=1}^{s-2}P_i + 3P_{s-1}\sum_{i,j=1}^{s-2}P_iP_j\bigg),\tag{A2}
\label{Eq:Qs}
\end{align}
where $\sum_s Q_s = 1-(1-pP_{\infty})^3$ is satisfied.

To calculate $P_s$ for a node in the first partition, the suppression rule must be taken into account. At first, the node has $0 \leq n \leq 3$ occupied links with probability $\binom{3}{n}p^n(1-p)^{3-n}$ as discussed in the main text. Then, the node occupies $n$ links in ascending order of neighboring cluster sizes, where each cluster size $i$ is randomly assigned according to $R_i$. To determine the cluster size $s$ of the node, we take 1 plus the maximum of the $n$ cluster sizes, similar to the calculation of $R_s$ and $Q_s$. 

As a result, $P_s$ is given by
\begin{widetext}
\begin{align}
P_{s}=&(1-p)^3\delta_{s1}+3p(1-p)^2 \bigg[ R_{s-1}^3+3R_{s-1}^2\bigg(1-\sum_{i=1}^{s-1}R_i\bigg) + 3R_{s-1}\bigg(1-\sum_{i=1}^{s-1}R_i\bigg)^2\bigg] \notag \\
       &+3p^2(1-p)\bigg[R_{s-1}^3+3R_{s-1}^2\bigg(1-\sum_{i=1}^{s-1}R_i\bigg)+3R_{s-1}^2\sum_{i=1}^{s-2}R_i
       + 6R_{s-1}\bigg(1-\sum_{i=1}^{s-1}R_i\bigg)\bigg(\sum_{j=1}^{s-2}R_j\bigg)\bigg] \notag \\
       &+p^3\bigg[R_{s-1}^3+3R_{s-1}^2\bigg(\sum_{i=1}^{s-2}R_i\bigg) + 3R_{s-1}\bigg(\sum_{i=1}^{s-2}R_i\bigg)^2\bigg],\tag{A3}
\label{Eq:Ps}
\end{align}
\end{widetext}
where the $(n+1)$th term from the left on the right-hand side represents the product of $\binom{3}{n}p^n(1-p)^{3-n}$ and the probability that the $(4-n)$th largest neighboring cluster size is $s$.
This equation satisfies Eq.~(\ref{eq:Pselfconsist}) for $z=3$. Using Eqs.~(\ref{Eq:Qs}) and (\ref{Eq:Ps}), we plot $sn_s = (P_s+Q_s)/2$ as theoretical guidelines in Fig.~\ref{Fig:Bipartite_Rewire_cdist}. 

\section*{Appendix B: discontinuous percolation with poisson and scale-free degree distributions}

We construct bipartite networks of size $N$ consisting of two partitions of $N/2$ nodes with Poisson and scale-free degree distributions. For the Poisson degree distribution, at each time step, one node is chosen uniformly at random from each partition, and a link is added between them if they are not already connected. For the scale-free degree distribution, at each time step, a node $i$ $(1 \leq i \leq N/2)$ is selected according to the probability $i^{-\alpha}/\sum_{i}i^{-\alpha}$, and another node $j$ $(N/2+1 \leq j \leq N)$ is selected according to the probability $(j-N/2)^{-\alpha}/\sum_j (j-N/2)^{-\alpha}$ for the parameter $\alpha \in [0, 1)$. A link is added between $i$ and $j$ if they are not already connected. For both degree distributions, $zN/2$ links are added so that the average degree of each node becomes $z$. As a result, the Poisson degree distribution is given by $P_k = (z^ke^{-z})/k!$, while the scale-free degree distribution follows
$P_k \propto k^{-\lambda}$, with $\lambda = 1+1/\alpha$ and satisfying $\sum_k kP_k = z$~\cite{UniversalLoad}.

\begin{figure}[t!]
\includegraphics[width=1.0\linewidth]{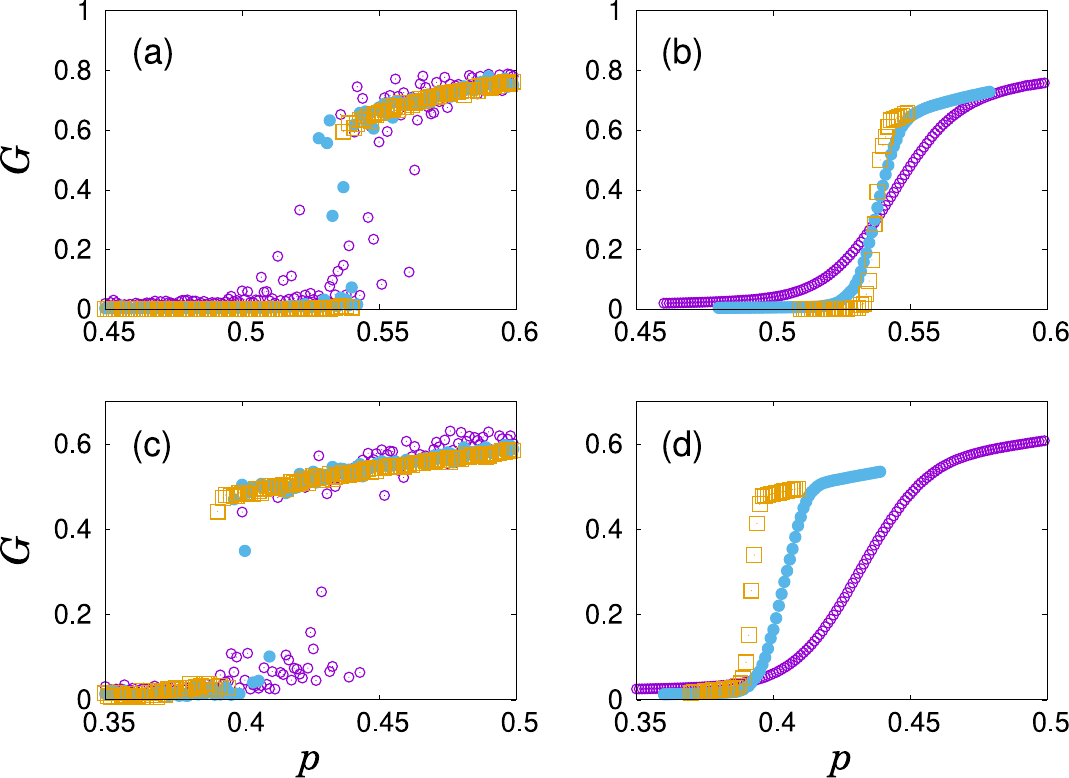}
\caption{(a,b) $G(p)$ of the Poisson degree distribution with $z=3$, shown for a single realization (a) and as an average over multiple realizations (b).
(c,d) $G(p)$ of the scale-free degree distribution with $z=3$ and $\lambda=2.5$, shown for a single realization (c) and as an average over multiple realizations (d).
In all four panels, $N/10^3 = 1$ $(\circ)$, $8$ $(\bullet)$, and $64$ $(\square)$ are used.}
\label{Fig:ER_SF_G}
\end{figure}

In Sec.~\ref{sec:model}, we introduced the model on a bipartite network with a constant degree $z$. Here, we apply the model to bipartite networks with the above Poisson and scale-free degree distributions. In Fig.~\ref{Fig:ER_SF_G}, we observe a DPT through the implementation of the model for both degree distributions. Specifically, $G(p)$ exhibits a discontinuous jump for a single realization, as shown in (a) and (c). When averaged over multiple realizations, $G(p)$ increases gradually but more rapidly with increasing $N$, as shown in (b) and (d), supporting the emergence of a DPT in the thermodynamic limit $N \rightarrow \infty$.

For the scale-free degree distribution, the threshold appears to decrease as $N$ increases, as shown in Fig.~\ref{Fig:ER_SF_G}(d).
By examining the behavior of the point of maximum slope of $G(p)$, similar to a previous study on DPTs in scale-free networks~\cite{hybridSF}, we estimate that the threshold converges to $0.38$ approximately, indicating that a DPT indeed emerges at a finite threshold.

\section*{Appendix C: comparison with previous Bethe lattice results}

\begin{table*}
\centering
\renewcommand\thetable{C1}
\caption{Theoretical estimates of the critical exponents $\tau$ and $\beta$}
\begin{tabular}{|c|c|c|}
\hline
               & Bethe lattice with $z-1=3$~\cite{yschosciadv2025}  & Bipartite network with $z=3$ \\ 
\hline
~~$\tau$~~    &       $2.29 \pm 0.01$  &   $3.0 \pm 0.1$   \\
\hline
~~$\beta$~~     &       1/2   & 1/2   \\
\hline
\end{tabular}
\label{Table:criticalexponents}
\end{table*}

In our previous study~\cite{yschosciadv2025}, a Bethe lattice with $z=4$ was considered. On the Bethe lattice, each node rewires its $z-1=3$ links in outward directions from the root to suppress the formation of large clusters, leading to a DPT in the emergence of the infinite spanning cluster.
Comparing this with the current results, in summary, each rewiring node rewires its three links to collaboratively suppress the formation of a macroscopic cluster on both the Bethe lattice and the bipartite network with $z=3$ considered in this study, leading to a DPT.

The theoretical values of $\beta$ and $\tau$ on the Bethe lattice were obtained in~\cite{yschosciadv2025}, as listed in Table~\ref{Table:criticalexponents}.
The theoretical estimate of $\tau$ for the bipartite network, derived in Sec.~\ref{sec:cdist}, is also included in Table~\ref{Table:criticalexponents}. We derive the theoretical value of $\beta$ for the bipartite network by examining the critical behavior $G(p)-G(p_c) \propto (p-p_c)^{\beta}$.

Using the relation $G(p)-G(p_c)=(P_{\infty}(p)+Q_{\infty}(p)-P_{\infty}(p_c)-Q_{\infty}(p_c))/2$ with $Q_{\infty}=1-(1-R_{\infty})^{3/2}$ and $P_{\infty}=(1-\sqrt{1-R_{\infty}})/p$, we derive the dominant behavior 
\begin{equation} 
G(p)-G(p_c) \propto \frac{\partial{R_{\infty}}}{\partial p}\biggr|_{p=p_c}(p-p_c) 
\tag{C1}
\end{equation} 
as $p \rightarrow p_c^+$, where $R_{\infty}(p)-R_{\infty}(p_c) \propto (p-p_c)^{\beta}$ as $p \rightarrow p_c^+$.

We analyze the dominant term of the physical solution $R_{\infty} > R_{\infty}(p_c)$ of $f(R_{\infty})=g(R_{\infty})$ as $p \rightarrow p_c^+$ [see Fig.~\ref{Fig:Bipartite_G_Theory}(a)]. Using the conditions $f(x)=g(x)$ and $f'(x)=g'(x)$ at $p=p_c$ with $x=R_{\infty}(p_c)$, the dominant terms of both sides of $f(R_{\infty})=g(R_{\infty})$ are obtained as
\begin{equation}
\frac{\partial f(R_{\infty}(p))}{\partial p}\biggr|_{p=p_c}(p-p_c) = \frac{1}{2}\frac{\partial^2 g}{\partial x^2}\biggr|_{x=R_{\infty}(p_c)}(R-R_{\infty}(p_c))^2,  
\end{equation}
which leads to $\beta = 1/2$. This result is presented in Table~\ref{Table:criticalexponents}.

\bibliography{references.bib}

\end{document}